\documentclass[preprint,trackchanges]{aastex62}

\newcommand{\speed}[1]{#1 km~s${}^{-1}$}
\newcommand{\accel}[1]{#1 km~s${}^{-2}$}
\newcommand{\nfig}[1]{Figure~\ref{#1}}
\newcommand{\tbl}[1]{Table~\ref{#1}}
\usepackage{ulem}
\usepackage{lineno}
\shorttitle{QFP wave train and accelerated electron}
\shortauthors{Duan et al.}

\begin{document}
\title{Homologous Accelerated Electron Beams, Quasi-periodic fast-propagating Wave and CME Observed in one Fan-spine Jet}
\correspondingauthor{Yuandeng Shen}
\email{ydshen@ynao.ac.cn}
\author[0000-0001-9491-699X]{Yadan Duan}
\affil{Yunnan Observatories, Chinese Academy of Sciences, Kunming, 650216, China}
\affil{State Key Laboratory of Space Weather, Chinese Academy of Sciences, Beijing 100190, China}
\affil{University of Chinese Academy of Sciences, Beijing 100049, China}
\author[0000-0001-9493-4418]{Yuandeng Shen}
\affil{Yunnan Observatories, Chinese Academy of Sciences, Kunming, 650216, China}
\affil{State Key Laboratory of Space Weather, Chinese Academy of Sciences, Beijing 100190, China}
\affil{University of Chinese Academy of Sciences, Beijing 100049, China}
\author{Xinping Zhou}
\affil{Yunnan Observatories, Chinese Academy of Sciences, Kunming, 650216, China}
\affil{University of Chinese Academy of Sciences, Beijing 100049, China}
\author{Zehao Tang}
\affil{Yunnan Observatories, Chinese Academy of Sciences, Kunming, 650216, China}
\affil{University of Chinese Academy of Sciences, Beijing 100049, China}
\author{Chengrui Zhou}
\affil{Yunnan Observatories, Chinese Academy of Sciences, Kunming, 650216, China}
\affil{University of Chinese Academy of Sciences, Beijing 100049, China}
\author{Song Tan}
\affil{Yunnan Observatories, Chinese Academy of Sciences, Kunming, 650216, China}
\affil{University of Chinese Academy of Sciences, Beijing 100049, China}

\begin{abstract}
Using imaging and radio multi-wavelength observations, we studied the origin of two homologous accelerated electron beams and a quasi-periodic fast-propagating (QFP) wave train associated with a solar jet on 2012 July 14. The jet occurred in a small-scale fan-spine magnetic system embedding in a large-scale pseudostreamer, which associated with a {\em GOES} C1.4 flare, a jet-like coronal mass ejection (CME), a type II radio burst, and a type III radio burst. During the initial stage, a QFP wave train and a fast moving on-disk radio source were detected in succession ahead of the jet along the outer spine of the fan-spine system. When the jet reached a height of about 1.3 solar radii, it underwent a bifurcation into two branches. Based on our analysis results, all the observed phenomena in association with the jet can be explained by using a fan-spine magnetic system. We propose that both the type III radio burst and the on-disk fast moving radio source were caused by the same physical process, i.e., the energetic electrons accelerated by the magnetic reconnection at the null point, and they were along the open field lines of the pseudostreamer and the closed outer spine of the fan-spine structure, respectively. Due to the bifurcation of the jet body, the lower branch along the closed outer spine of the fan-spine structure fell back to the solar surface, while the upper branch along the open field lines of the pseudostreamer caused the jet-like CME in the outer corona. 
\end{abstract}


\section{Introduction} \label{sec:intro}
Solar jets represent collimated transient plasma ejections along magnetic field lines in the solar atmosphere. These ubiquitous transient events are frequently detected at H$\alpha$, ultraviolet, extreme ultraviolet (EUV), and X-ray wavebands, and they have received great attention due to their potential role in providing significant mass and energy input into the upper atmosphere and the solar wind \citep{2021RSPSA.47700217S}. 
A consensus has been reached in recent years that the formation of solar jets is closely related to the magnetic reconnection process. Observations indicate that solar jets often occur in mixed-polarity regions, and both flux emergence and flux cancellation in which are important for triggering solar jets \citep[e.g.,][]{2007A&A...469..331J,2017ApJ...844..131P,2019ApJ...887..220Y,2021ApJ...912L..15T,2019ApJ...887..239Y}. High spatiotemporal resolution observations have revealed that solar jets are always accompanied by flaring bases manifesting as brighten points (or patches), which included key information about their formation. Focusing on the morphological evolution in time using X-ray observations, \cite{2010ApJ...720..757M} firstly proposed that one-third of coronal jets could be recognized as blowout jets whose eruption include the complete ejection of the cool and shear core fields and resemble the eruptions of large-scale coronal mass ejections (CMEs). \cite{2012ApJ...745..164S} directly observed in H$\alpha$ images that the cool ejecting shear core is actually a mini-filament, whose eruption directly forms the cool component in the subsequent jet body. Occasionally, some strong mini-filament-driven blowout jets can result in narrow white-light jets (narrow CMEs) \citep[e.g.,][]{2011ApJ...738L..20H,2012ApJ...745..164S,2019ApJ...881..132D}. In recent years, many high resolution EUV observations with the aid of magnetic extrapolation have revealed that the basic three-dimensional magnetic topology of solar jets is the so-called fan-spine topology, which consists of a dome-shaped fan, inner and outer spines passing through a three-dimensional null point \citep[e.g.,][]{2009ApJ...691...61P,2009ApJ...704..485T,2009ApJ...700..559M,2012ApJ...760..101W,2019ApJ...885L..11S,2020ApJ...898..101Y,2021ApJ...923...45Z}. Such a magnetic configuration often builds up as a magnetic polarity intrudes into a dominate opposite-polarity region. Sometimes, small-scale filaments or sheared core fields form and erupt under the fan dome due to continual photospheric rotation/shear motions and flux cancellation  \citep[e.g.,][]{2017ApJ...844..131P,2018SoPh..293...93C,2020ApJ...902....8C}, which subsequently results in the null point reconnection and the formation of solar jets.

Many numerical simulations conduct parametric studies on solar jets under fan-spine magnetic system  \citep[e.g.,][]{2015A&A...573A.130P,2016A&A...596A..36P,2017Natur.544..452W}, and successfully reproduce their main observed properties, including plasma ejection, helical jet morphology, and so on. In particular, several recent simulation works highlight that the null point in fan-spine magnetic systems is a favorable site for magnetic reconnection and related electrons acceleration \citep[e.g.,][]{2013ApJ...771...82M,2019ApJ...884..143M}. Observationally, type III radio bursts are useful to diagnose the outward propagating accelerated electrons, associated with solar jets, because the open field lines of the latter provide a natural path for the former to escape into the interplanetary space at semi-relativistic speeds \citep[e.g.,][]{2017ApJ...851...67S,2017ApJ...835...35H,2021A&A...647A.113Z}. For confined fan-spine jets where their outer spines are closed field lines rooting on the solar surface \citep{2021RSPSA.47700217S}, the accelerated electron beams should be trapped in the closed outer spine field lines. So far, direct observation of the electron acceleration site and transport process in solar jets is still not an easy task, and the acceleration mechanism is also unclear, although various candidate mechanisms involving the magnetic reconnection process have been proposed in previous articles \citep[e.g.,][]{2011ApJ...742...82K,2012ApJ...754....9G}. Recently, \cite{2018ApJ...866...62C} found that the electron beams were originated from an extremely compact region located behind the erupting jet spire but above the closed arcades, coinciding with the site of the null point. This result suggests that the acceleration of the observed electron beams probably took place in the null point reconnection.

Quasi-periodic fast-propagating (QFP) wave train are one new types of EUV waves first observed by SDO/AIA \citep{2011ApJ...736L..13L}. Compared with the ``single-pulse" EUV waves or Morton waves that freely propagate in very wider angular extents to greater distances across the solar disk, QFP waves are usually characterized as a series of arc-shaped, multi-wavefront of EUV emission variation and more often travel along funnel-shaped coronal loops with a high speed of \speed{500$-$20000} \citep{2014SoPh..289.3233L,2018ApJ...853....1S}. These unique observed properties of QFP waves are thought to be strong evidence of quasi-periodic fast-mode magnetosonic waves \citep{2011ApJ...740L..33O,2014SoPh..289.3233L}. Generally, the generation of QFP wave train in the corona should be diversified, but it is most likely in association to mechanisms of pulsed energy release in magnetic reconnections and the dispersive evolution of an impulsively broadband perturbation \citep{2021arXiv211214959S,2021SoPh..296..169Z}. The leakage of photospheric and chromospheric oscillation into the corona \citep{2012ApJ...753...53S}, the untwisting motions of helical threads consisting of filaments, and CMEs are both thought as candidate drivers of QFP wave train \citep{2019ApJ...873...22S,2019ApJ...871L...2M}. Of particular note that in solar flares triggering in fan-spine magnetic systems, the simultaneous occurrence of periodic radio bursts and QFP wave train are evidenced \citep{2013SoPh..283..473M,2017ApJ...844..149K}. This naturally implies that null point reconnection in fan-spine systems may also launch QFP wave train. However, what is the detailed physical relationship between null point reconnection and the associated QFP wave train is still an open question. 

In this paper, we study a solar EUV jet on 2012 July 14, which occurred in a fan-spine magnetic system hosted by a large-scale pseudostreamer. Various phenomena including a QFP wave train, a fast moving on-disk radio source, a narrow CME, a type II radio burst and a type III radio burst are observed to be associated with the jet. Based on our multi-wavelength observational analysis, we attempt to explain all these jet-related phenomena within the framework of fan-spine magnetic topology.

\section{RESULTS} \label{sec:OBS}
\subsection{Pre-eruption Magnetic Condition and the Eruption}
Figure 1 presents the overview of the solar jet of our interest. The Potential Field Source Surface (PFSS; \citet{2003SoPh..212..165S}) technique provided by the standard SolarSoftWare package is used to analysis the global three-dimensional (3D) magnetic field configuration of the eruptive region. The extrapolated results in \nfig{fig1} (a) and (d) are obtained from the same global PFSS extrapolation, whose bottom boundary corresponds to a synoptic HMI magnetic magnetogram at 06:00 UT. It is clear that the jet took place at the south edge of a larger-scale pseudostreamer. In AIA 304 \AA \ imaging observation, this jet had a relatively broad spire and its triggering process involved the eruption of a small filament (see \nfig{fig1} (b)). This suggests that the observed jet was a blowout jet as proposed in \citet{2010ApJ...720..757M}. The eruption source region was composed of a positive polarity surrounded by negative background magnetic polarities, and with a mini-filament lying along the west polarity inversion line (see  \nfig{fig1}(c). Such a magnetic condition provides a desirable condition to breed a fan-spine topology.
As revealed by the closed-in extrapolated result in \nfig{fig1} (d), the fan-spine structure was embedded in the large-scale pseudostreamer, together comprising a unique coupling magnetic system. To better charaterize the three-dimensional magnetic field of the small-scale fan-spine structure, we took the line-of-sight HMI magnetogram (xrange=[475,700], yrange=[63,360]) at 08:22 UT as the bottom boundary and performed a new local potential field extrapolation with NLFFF code provided by SolarSoftWare (SSW).
The local extrapolation is performed within the cubic box of 376$\times$496$\times$201 grid points with $\bigtriangleup$x = $\bigtriangleup$y = $\bigtriangleup$ z= 0\arcsec.5, and the result projected onto AIA 1600 \AA\ image (see \nfig{fig1} (d1)). In which, the red lines outline its skeleton and the null-point position is found to appeared around 5.05 Mm above the photosphere (as marked by the white arrow). During the eruption of the solar jet, the footpoints of the fan surface well coincides with an AIA 1600 \AA \ circular ribbon in the \nfig{fig1} (d1). This is well agree with the three-dimensional reconnection nature in a fan-spine magnetic configuration \citep[e.g.,][]{2016A&A...591A.141J}. 
Of particular note that the outer spine lines of the small-scale fan-spine structure in \nfig{fig1} (d) bifurcate themselves into two part: green lines are open to the interplanetary space, while white lines are confined to a remote region.

\nfig{fig2} shows the mini-filament eruption and the flare ribbons in the eruption source region with AIA 304 and 1600 \AA\ images, respectively. The eruption of the mini-filament underwent a short slow rising phase from about 09:02 UT to 09:07 UT, after that it accelerated and erupted violently (see \nfig{fig2}(a1--a4)). A time-distance made along the white arrow in \nfig{fig2}(a1) is plotted in \nfig{fig2}(c), which clearly displays the slow and fast eruption phase of the mini-filament. By tracking specific observed features of the erupting filament (the dark streak) and the subsequent jet plasma ejection (the bright streak) in \nfig{fig2}(c), we obtain that the projected velocity during the two phases was about  {\speed{6.7 ± 1.8}} and  {\speed{37.9± 8.1}} , respectively. Their speeds (errors) are given by one average value (standard derivation) of five different measurements.
In addition, we also obtain the acceleration of the filament eruption was about \accel{0.22} by fitting the filament trajectory with a quadratic function. The eruption was accompanied by a {\em GOES} C1.4 flare whose start, peak, and end times were at 09:08 UT, 09:13 UT, and 09:30 UT, respectively (see \nfig{fig5} (a1)). In the AIA 1600 \AA\ images, one can clearly identify the formation of an inner bright patch surrounded by a circular ribbon that respectively manifested the footpoint locations of the inner spine and the fan structure of the fan-spine system (also see \nfig{fig1}(d1)). At 09:15 UT when the circular ribbon faded out, two elongated, conjugated flare ribbons appeared at both sides of the mini-filament  (see \nfig{fig2}(b1--b4)). According to the initial magnetic field environment of the present event, these eruption features and temporal relations among them suggest that the rising mini-filament firstly triggered the magnetic reconnection around the null point of the fan-spine system, then the electron beams was accelerated in the reconnection streamed downward along the field lines of the inner spine and fan structure and impacted upon the low-altitude atmosphere to cause the inner bright patch and the circular ribbon, respectively \citep[e.g.,][]{2021APJ...923..163}. In the meantime, the null point reconnection also successively removes the confinement field of the slow-rising mini-filament, such positive feedback finally leads to the violent eruption of the mini-filament and the formation of two elongated ribbons.

\subsection{QFP Wave train and CME}
At the launch of the EUV jet, an EUV wave train was detected ahead of the jet spire and eventually propagated to the north solar limb. This wave train can be best identified in AIA 171 \AA\ passbands. More interestingly, the wave train demonstrated as a multiple wavefront nature and remain an arc shape during their propagation (see running difference images in \nfig{fig3} (c1)--(c4)) and its animation). These observed features are distinguished from those freely propagating EUV waves and Morton wave, thus better qualifying a QFP wave train. To better describe the temporal and spatial relationship between the wave train and the EUV jet, a space-time plot is made along the blue dotted lines in \nfig{fig3} (a). In \nfig{fig3} (b), it can be seen that the wave train first appeared at about 09:14 UT, which is around 5 minutes after the launch of the EUV jet (09:09 UT). Moreover, the wave train was observed at a spatial distance of about 400 Mm far away from the flare source region. The velocity of the EUV jet and the wave train were \speed{360} and \speed{538}, respectively.
The kinematic of the wave train is measured along a cut of a sector in  \nfig{fig3} (a). This sector is a function of distance measured from the wave along the longitudinal great circle by averaging pixels in the latitudinal direction, correcting the inherent sphericity of the solar surface \citep[e.g.,][]{2010ApJ...723L..53L,2020ApJ...894...30W}. The corrected velocity of the wave train is derived at \speed{719} and it propagated up to a distance of 250 Mm from 09:14 UT to 09:20 UT. For the EUV jet, its projection speed of the jet on the solar disk is derived around \speed{360} based on the AIA observations. To minimum its projection effect, we re-estimate the propagation velocity of the EUV jet in the {\em STEREO}-A view along the dashed blue curve in \nfig{fig4} (a1).  As a result, we find that the speed of the EUV jet was about \speed{425}. These observed features indicate that the generation of QFP wave train is likely related to the launch of the EUV jet.

From the {\em STEREO-A} perspective, we found that the jet experienced a bifurcation phenomenon when it reached a height of about 1.3 solar radii above the solar surface (see the arrows in \nfig{fig4}(a2) and (d)). 
One possibility for the jet bifurcation is that the jet material simultaneously entered open and closed magnetic field line after the initial null-point reconnection. We notice that a similar jet dynamic process was reported in the recent simulation work from \citet{2021ApJ...909...54W}. By projecting the global extrapolated field lines from PFSS model onto STEREO-A EUVI 304 \AA \ image (see \nfig{fig4} (d)), we can found that there are indeed two different system of magnetic field lines, i.e., the green open and the white closed lines. As this jet bifurcation proceeded, a portion of jet material was confined by the closed filed lines and finally impacted on the solar surface and resulted in a EUV brightening, which might be located near the remote footpoint of the outer spine of the fan-spine system. The brightening reached the remote solar surface was at 09:25 UT (see \nfig{fig4}(a4)). Assuming that it launched with the EUV jet, we can roughly estimate  the propagation speed of the EUV brightening to be about \speed{906}, which seems to be comparative to the projected speed of QFP wave train (\speed{719}). On the other hand, the rest of jet material ejected to the high corona along open field lines of the pseudostreamer, as denoted by black arrows in EUVI 304 \AA\ images (see \nfig{fig4}(a3)). The lower and the upper branches projection speeds were about \speed{425 and 328}, respectively. Caveat that these rough speeds suffer from inevitable observational errors because only a few EUVI data points are available for our speed measurement. 
Similar to previous observations \citep[e.g.,][]{1998ApJ...508..899W,2009SoPh..259...87N,2019ApJ...881..132D,2020ApJ...901...94J,2020ApJ...900..158Y}, as the jet flow ejected along the open flux, a narrow jet-like CME was detected in the FOV of {\em STEREO-A} COR1 white-light observations (see \nfig{fig4} (b)) and LASCO/C2 (c), with an angular width of about $15^{\circ}$. It is measured that speeds of the CMEs observed by {\em STEREO-A} COR1 and LASCO/C2 were about \speed{540 and 561}, respectively.

\subsection{Radio Signatures}
At the very beginning of the fan-spine jet (during 09:08:51$-$09:09:56 UT), a fast moving radio source was detected by the full-disk imaging radio observation from NRH at 150 MHz. As shown in \nfig{fig5} (b1)-(b3), the radio source appeared firstly near the jet base at 09:08:51 UT, and then it rapidly moved to the northwest of the solar disk like the observed jet and the QFP wave train (see the right column of Figure 5 and Figure 3(d)). Combined with the global extrapolated coronal field fields from the PFSS model, one can notice that the propagating path of the moving radio sources most likely along the close coronal field lines (see \nfig{fig3} (d)). Different from the relative stationary radio sources that observed during large-scale solar eruptions \citep[e.g.][]{2020FrASS...7...79C,2021ApJ...911...33C}, the fast moving on-disk radio source had a speed of in the range of \speed{4146 to 8414}  (i.e., $\sim$ 0.0138$c-$0.028$c$), and a height of about 1.46 solar radii
above the solar surface.  Nearly at the same time (09:08 UT), an interplanetary type III burst was also detected by the NDA and {\em WIND/WAVES}, (see Figure 5 (a2) and (a3).), which indicates that the appearance of accelerated electrons along the open field lines of the pseudostreamer. Based on the coronal plasma density model presented by \citet{1999ApJ...523..812S}, we derived the speed of the type \uppercase\expandafter{\romannumeral3} radio burst was in the range of \speed{6934$-$101385} (i.e., 0.01$c-$0.3$c$).
These observational characteristics and their intimate temporal relationships might suggest that both the type III radio burst and the simultaneous on-disk fast moving radio source were both caused by the accelerated electrons produced in the null point reconnection (09:07$-$09:09 UT), but they were along different magnetic field lines. Specifically, those electrons along the open field lines of the pseudostreamer resulted in the type \uppercase\expandafter{\romannumeral3} radio burst, while those along the closed outer spine of the fan-spine system caused the on-disk fast moving radio source.

In addition, a type  \uppercase\expandafter{\romannumeral2} radio burst was also observed right after the type \uppercase\expandafter{\romannumeral3} radio burst at about 09:11 UT (see \nfig{fig5} (a2)). Generally, type \uppercase\expandafter{\romannumeral2} radio burst is thought to be caused by interplanetary shocks. Based on the coronal plasma density model presented by \citet{1999ApJ...523..812S}, we estimated that the height and the speed of the shock were about 1.57$-$1.68 solar radii and \speed{286}, respectively. In addition, by measuring the band-split distance of the type  \uppercase\expandafter{\romannumeral2} radio burst, we obtained that the density compression ratio of the shock and the local Alfv\'{e}n speed was about 1.32 and \speed{240$-$260}, respectively.
\section{Discussion and Conclusion} \label{sec:summ}
\begin{figure}   
\centerline{\includegraphics[width=0.9\textwidth]{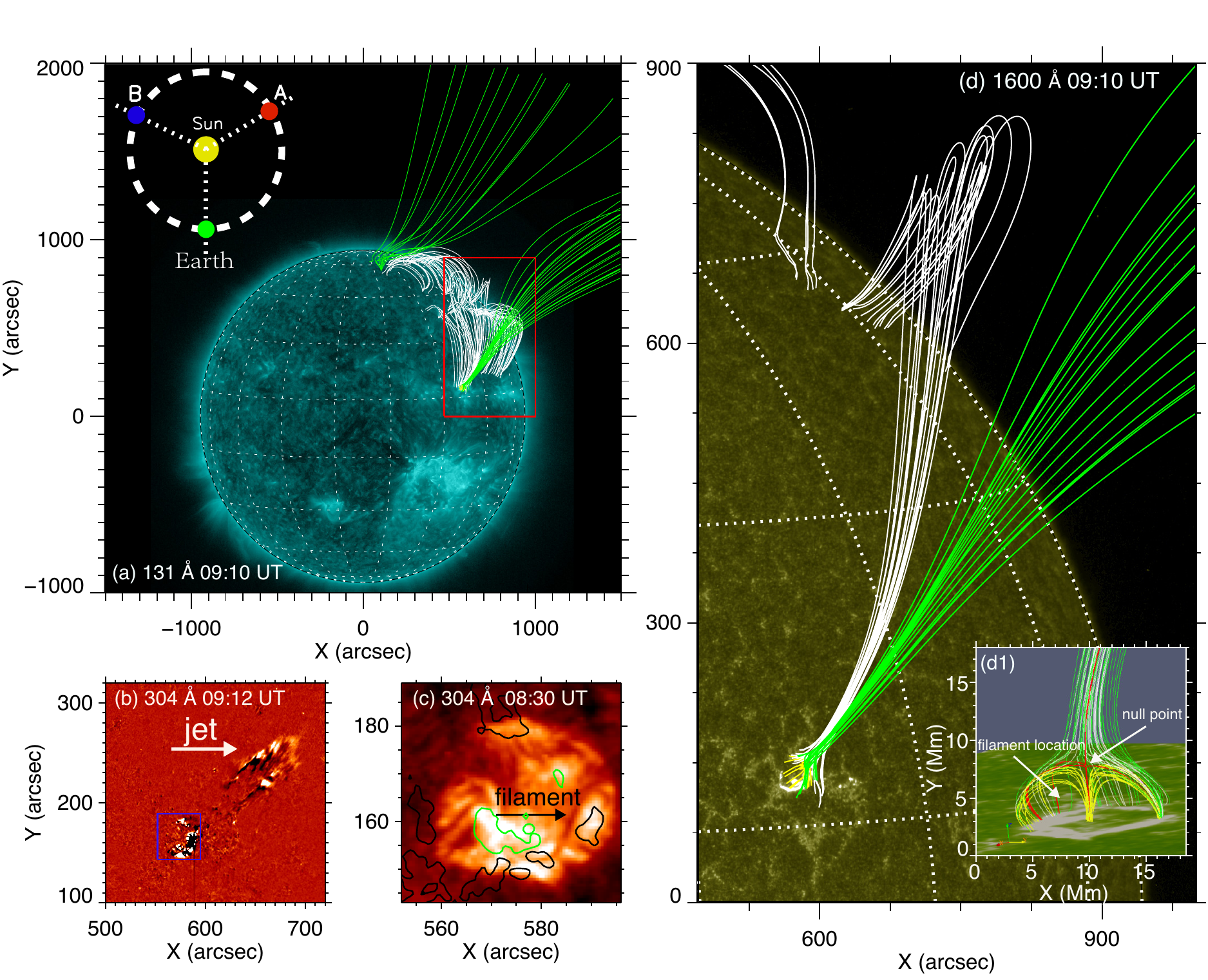}}
\caption{Panel (a) shows the overview of the source region, in which the erupted region of the AIA 131 \AA\ image was overlaid the extrapolated PFSS field lines. The position of the Earth and two satellites of STEREO relative to the Sun as shown in the upper left corner, the separation angle between SDO and STEREO Ahead was about 120 degrees. The red box indicates the field of view (FOV) of the panel (d). Panel (b) : the  {AIA} 304 \AA\ running-difference image displays the morphology of the jet, and the blue box represents the FOV of the panel (c).  {Panel (c) : the AIA 304 \AA\ image displays an S-shaped small filament at the base of the jet before it ejects}. The black and green contours represent negative and positive magnetic fields of the HMI LOS magnetogram scaled at ${\pm}$ 100G, respectively.  {Panel (d) : the reconfigured magnetic configuration of the pseudostreamer from the global PFSS model, which was projected onto AIA 1600 \AA\ image and a smaller-scale fan-spine topology existed at its south edge. The inset image (d1) gives a closer-up 3D view of the fan-spine topology at the south edge of the pseudostreamer, which was reconfigured by another local potential field extrapolation.}
In panel (d1), the skeleton is outlined by the red curves, and the position of the null point is indicated by a white arrow. In the extrapolation images in  {panels (a), (d) and (d1)}, the white and yellow lines respectively show the outer (inter) fan structures, while the green ones represent the south part of the pseudostreamer.  {Note that the green lines in panel (d1) may not be fully open, since the local extrapolation is performed with a limited height ($\sim$ 73 Mm)}.}
\label{fig1}
\end{figure}

\begin{figure}    
\centerline{\includegraphics[width=0.9\textwidth]{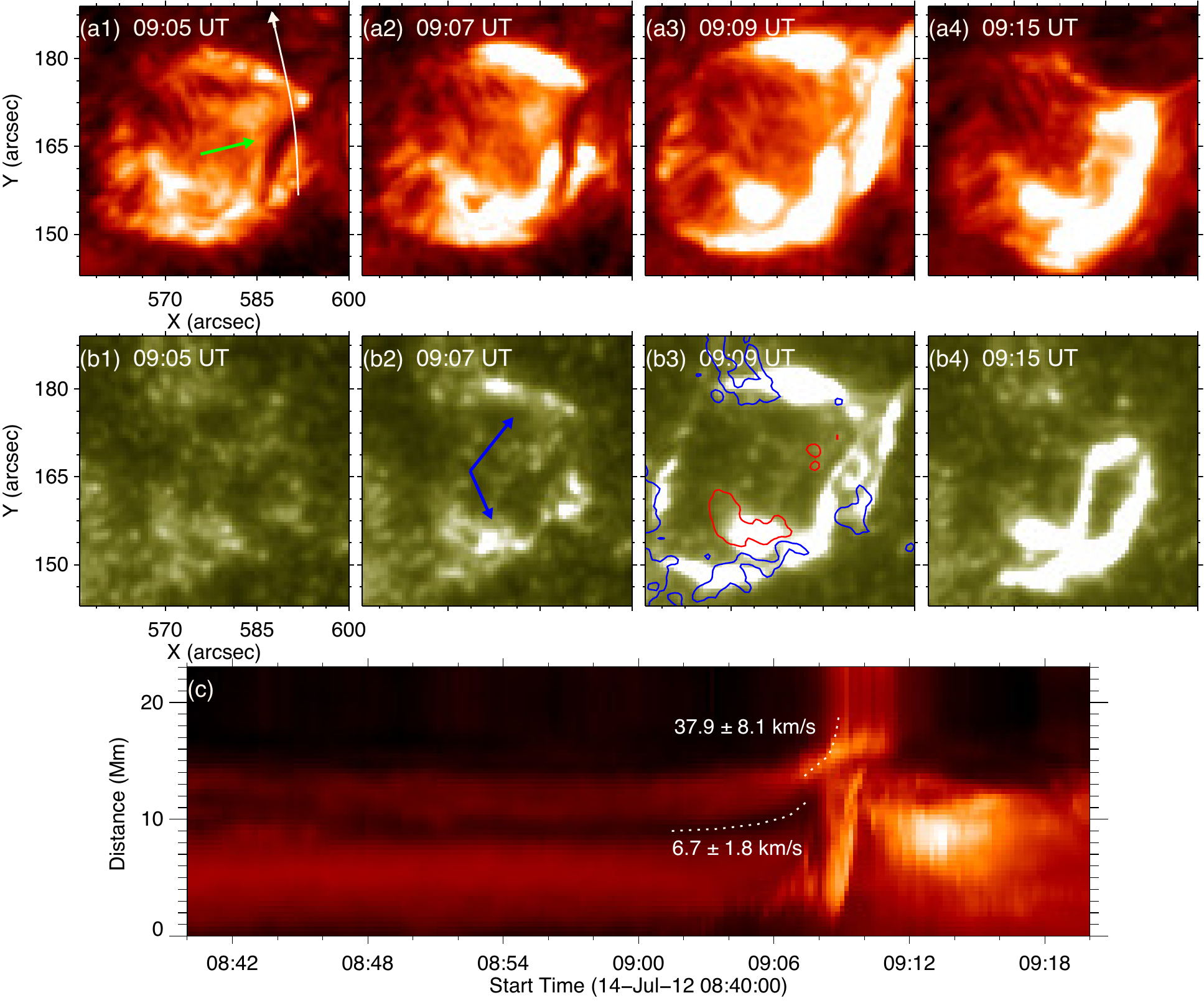}}
\caption{Panel (a1)-(a4): AIA 304 \AA\ sequence images show the eruptive process of the mini-filament.The green arrow points to the filament in panel (a1). Panel (b1)-(b4): time sequence of AIA 1600 \AA\ images demonstrate the flare ribbon variations during the eruptive process. The blue arrows in panel (b2) indicate a semi-circular flaring feature. The blue and red contours in panel (b3) are negative and positive magnetic fields of the HMI LOS magnetogram scaled at ${\pm}$ 100G, respectively. Panel (c): The time-distance plot along the white arrow in panel (a1).  {In which, the dark horizontal streak around 10 Mm corresponds to the rising filament body, while bright vertical streaks near 09:08 UT correspond to plasma ejections in the solar jet. These two features are used to roughly derive the projected velocity in the dynamic analysis in Section 2.1.}}
\label{fig2}
\end{figure}

\begin{figure}    
\centerline{\includegraphics[width=1\textwidth,clip=]{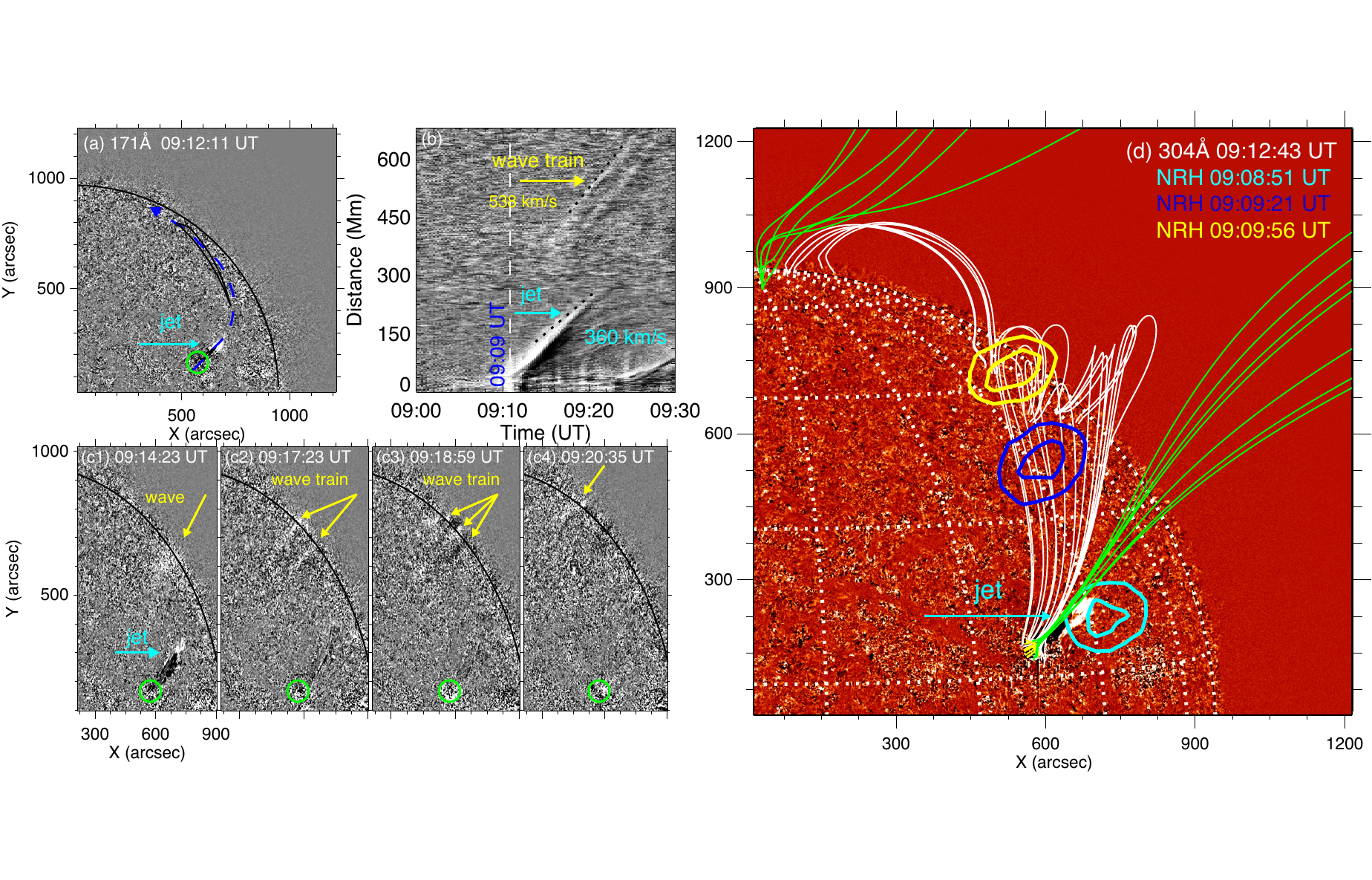}}
\caption{Panel (b) shows the time-distance plot along the blue curve arrow in panel (a), and the white dotted line denotes the start time of the jet. The black fan shape in panel (a) denotes the sector we used to measure the modified propagation velocity of the wave train. Panel (c1)-(c4): time sequence of AIA 171 \AA\ running difference images show the relationship between the jet and the QFP wave train. The yellow (cyan) arrows denote the wave train (jet), and the green circles indicate the eruptive source region. Panel (d): AIA 304 \AA\ image is overlaid with the extrapolated PFSS field lines, and the three different color contours represent NRH radio sources at different times (also see \nfig{fig5}(b1)--(b3)).  An animation of panel (a) is available. The animation covers 09:05 UT$–$09:20 UT with a 12 s cadence. The animation duration is 1 s.
}
\label{fig3}
\end{figure}

\begin{figure}    
\centerline{\includegraphics[width=1\textwidth,clip=]{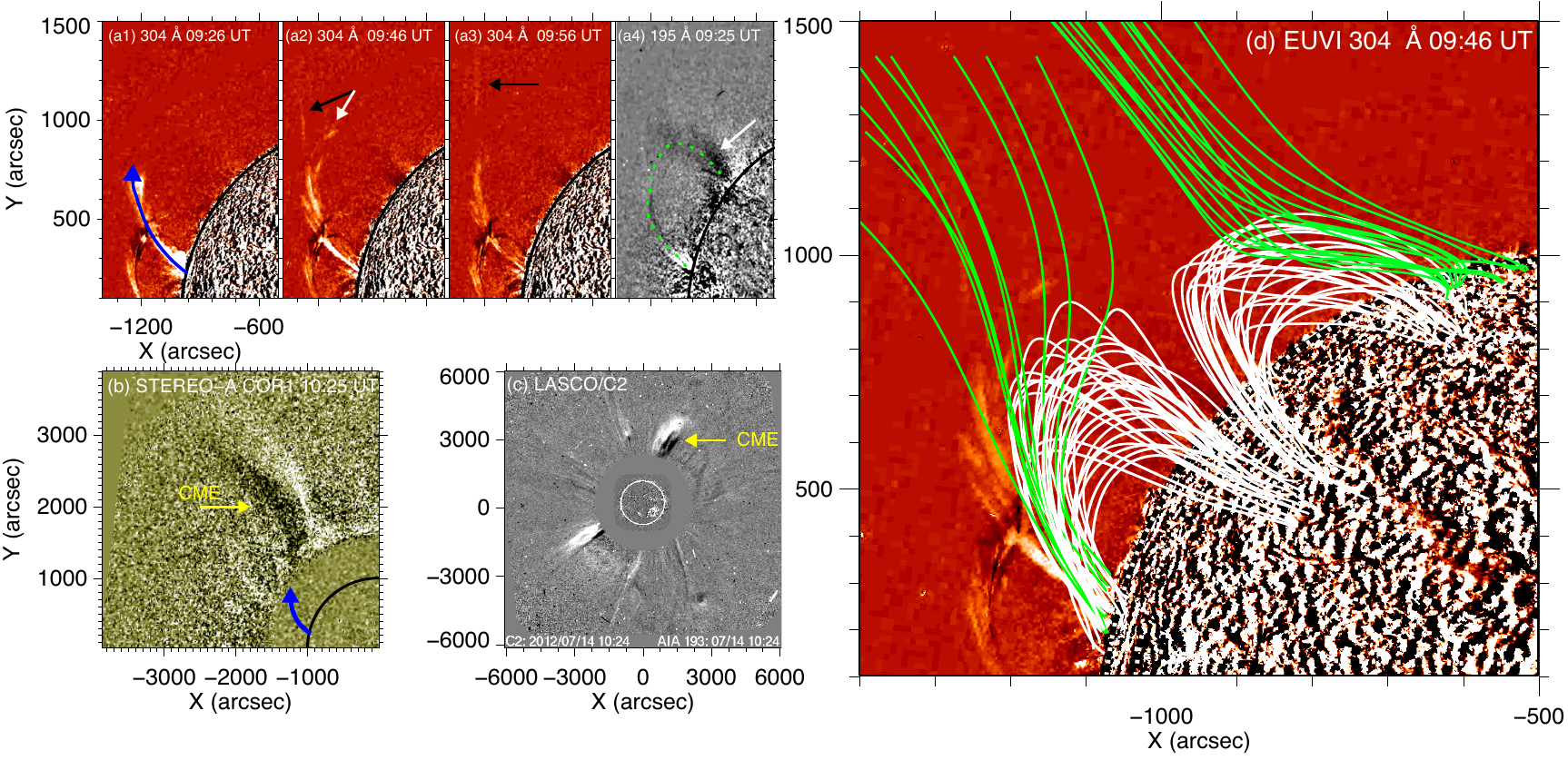}}
\caption{Panel (a1)-(a3) are time sequence of {\em STEREO}-A 304 \AA\ running difference images, and panel (a4) is the {\em STEREO}-A 195 \AA\ running difference image. In panel (a1)--(a4), the black and white arrows mark the traveling ejected material along different magnetic field lines.  {The green dotted line represent the propagating trajectory of the EUV brightening, with a travel distance about 816 Mm.} Panel (b) and (c): {\em STEREO}-A COR1 and LASCO/C2 images show a narrow CME. The blue arrow marks the ejected direction of the jet in panel (b). Panel (d): the extrapolated PFSS field lines overlay in the EUVI 304 \AA\ image, and the white (green) lines represent closed (open) magnetic fields.}
\label{fig4}
\end{figure}

\begin{figure}    
\centerline{\includegraphics[width=0.6\textwidth,clip=]{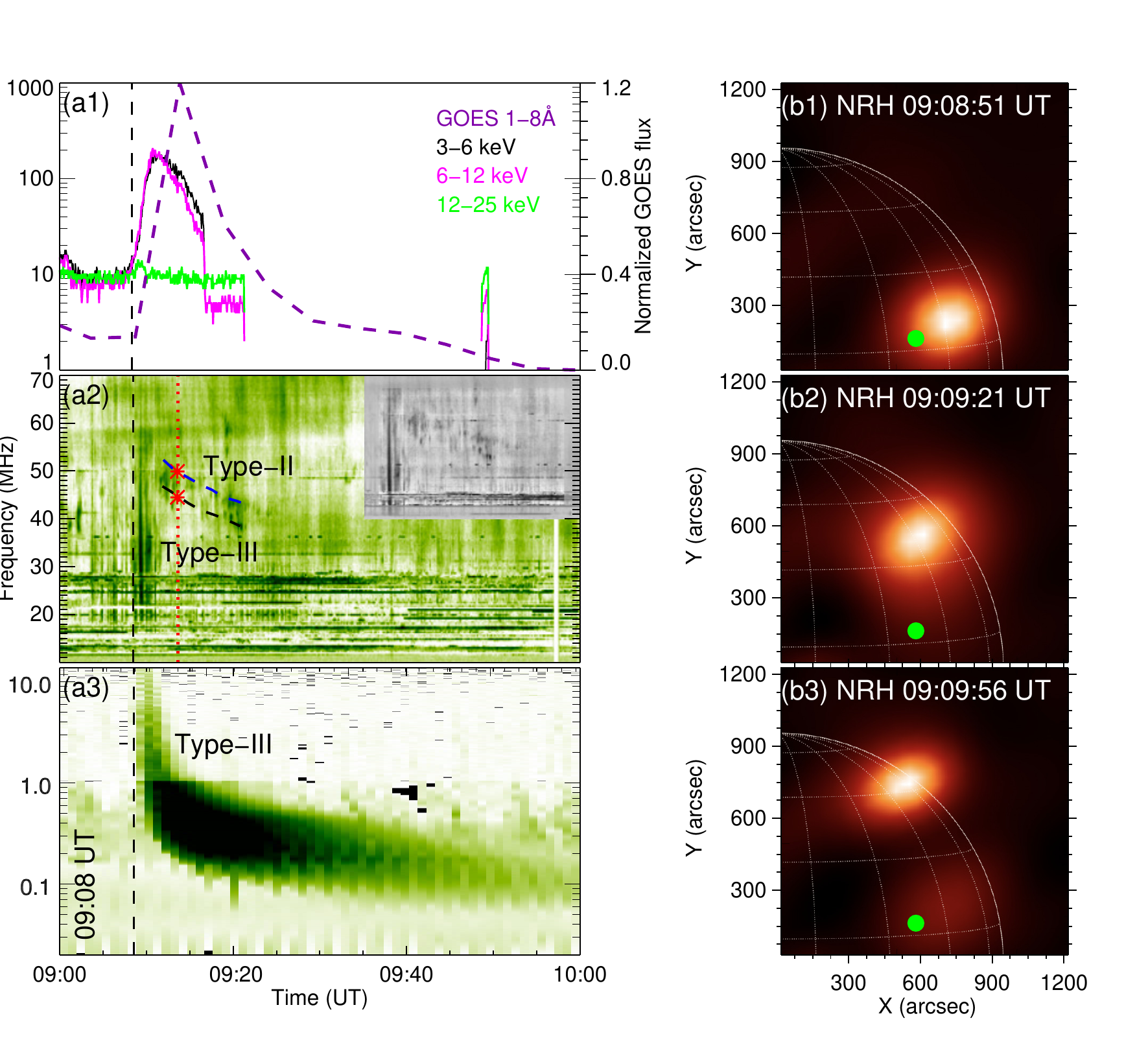}}
\caption{Panel (a1) shows the light curves of GOES 1-8 \AA\ and RHESSI in multiple energy channels from 09:00 UT to 10:00 UT. Panel (a2)-(a3): the radio dynamic spectrum from the Nançay Decametric Array (10-70 MHz) \citep[NDA;][]{1980Icar...43..399B} and RAD1 (RAD2) onboard WIND/WAVES(0.1-10 MHz)\citep{1995SSRv...71..231B} show the type II radio burst and the type III radio burst. The two asterisks in (a2) mark the frequencies used to calculate the compression ratio. Panel (b1)-(b3) show the time sequence of the Nançay Radioheliograph \citep[NRH;][]{1997LNP...483..192K} radio source; the green dots represent the eruptive source region.}
\label{fig5}
\end{figure}

The chronological order and the associated physical parameters of the eruption features described thereinbefore are listed in \tbl{tab1}, which can help us to make clear the temporal and spatial relationships among them. In combination with the extrapolated coronal magnetic field as presented in \nfig{fig1}, we propose that all the observed eruption features can be explained by using a fan-spine magnetic system embedding in a large-scale pseudostreamer. The following is the physical picture of the eruption. The low-lying mini-filament below the fan structure firstly starts to rise slowly due to some reason (for example, magnetic emergence or cancellations), which will squeeze the overlying magnetic fields and lead to the formation of a current sheet around the coronal null point \citep{2017Natur.544..452W}. The magnetic reconnection within the current sheet not only removes the confining magnetic field of the mini-filament but also accelerates electrons \citep{2018ApJ...866...62C}. Subsequently, the downward accelerated electrons result in the formation of the inner bright patch and the circular ribbon (see \nfig{fig2} (b3)), while those upward ones along the open field lines of the pseudostreamer and the closed field lines of the outer spine of the fan-spine system result in the formation of the observed type  \uppercase\expandafter{\romannumeral3} radio burst and the on-disk fast moving radio source. Due to the continuous reduction of the confinement magnetic field overlying the mini-filament, the slow rising mini-filament losses of its equilibrium completely and therefore erupts violently after the appearance of the type \uppercase\expandafter{\romannumeral3} radio burst, the on-disk fast moving radio source, the inner bright patch, and the circular ribbon. The eruption of the mini-filament can be interpreted by classical filament eruption models \citep{2011LRSP....8....6S}, in which the reconnection between the two legs of the confining magnetic field lines underneath the filament can naturally result in the two observed elongated ribbons. Due to the bifurcation of the jet body at a higher altitude, the upper and lower branches respectively caused the formation of the narrow CME in the outer corona and the QFP wave train along the closed outer spine of the fan-spine system.

Soar jets are frequently observed in the special fan-spine magnetic systems and are closely associated with the magnetic reconnection process. For some large-scale fan-spine systems whose outer spines extend into the interplanetary, they naturally provide an avenue for energetic electrons accelerated in the magnetic reconnection process to escape into the interplanetary space and therefore lead to the formation of type \uppercase\expandafter{\romannumeral3} radio bursts in radio spectrums \citep{2013ApJ...771...82M,2017ApJ...844..149K,2019ApJ...884..143M,2021RSPSA.47700217S}. Many previous observations also evidenced that type \uppercase\expandafter{\romannumeral3} radio bursts and impulsive energetic electron events are typically associated with solar jets \citep[e.g.,][]{2012ApJ...759...69W,2016ApJ...833...87C,2017ApJ...835...35H,2018ApJ...866...62C,2019A&A...632A.108M,2021A&A...647A.113Z}. For solar jets occurred in relatively small-scale fan-spine systems in which the remote footpoints of the outer spines can be identified on the solar surface, to the best of our knowledge, so far the signatures of accelerated electrons have not yet been detected in the closed outer spines. Therefore, the present observation of the on-disk moving radio source presented the first observation of accelerated electrons moving along the closed outer spine of a small-scale fan-spine magnetic system. In addition, based on the close temporal relationship between the type \uppercase\expandafter{\romannumeral3} radio burst and the on-disk fast moving radio source, we proposed that they were homogenous electrons accelerated by the same physical process (i.e., the null point reconnection) but propagated along the open field lines of the pseudostreamer and the closed outer spine of the fan-spine system, respectively. The observation of the homogenous accelerated electrons along with different magnetic field systems also suggests that the open field lines of the pseudostreamer surrounding the null point of the fan-spine system also took part in the null point reconnection beside the field lines of the fan-spine system itself. More similar observations and theoretical studies are desirable in the future to clarify the true physical process proposed in the present study.

The generation of the QFP wave train previously have been attributed to the dispersive evolution of a broadband perturbation and the pulsed energy release in association with the magnetic reconnection process \citep{2014SoPh..289.3233L,2021arXiv211214959S}. 
In the present event, the first appearing position of the QFP wave train was near the solar limb and was greater than 400 Mm from the flare epicenter, which is much longer than those reported in previous studies \citep{2021arXiv211214959S}. 
It remains unclear why such an obvious wave train do not appear at an early time, but we suggest that the detection of the QFP wave train might need to satisfy some specific conditions, such as  strong emission amplitudes of wave fronts, a right observed direction, and even a good enough instrument sensitivity.
In addition, the period of the wave train is also different from that of the flare.
Therefore, it is hard to directly account for the generation of the QFP wave train in terms of the two mechanisms for QFP wave train. Interestingly, we found that the QFP wave train appeared ahead of the jet  {spire} and propagated along the same trajectory with the jet \citep{2018ApJ...861..105S}.  More importantly, the first appearing time of the wave train delayed the start of the jet about 5 minutes, and its maximum speed (\speed{719}) is comparable to the typical coronal Alfv$\acute{e}$n speed, which was much faster than the jet  {spire}.  
These observed features indicate that the generation of QFP wave train is more likely related to the launch of the solar jet. Recently, \cite{2018ApJ...861..105S} firstly reported that single-pulsed EUV waves can be directly launched by coronal jets in coronal loops.  Such an excitation mechanism is similar to the generation of piston shocks in a one-dimensional tube, in which the jet acted as the driver of the wave train that can propagate faster than its driver \citep{2015LRSP...12....3W, 2018ApJ...856...24Y,2021ApJ...909....2M}. According to the piston shock scenario, our present QFP wave train could be driven by the jet ejection, in which two QFP wave fronts might related to unresolved discrete fine structures consisting of the jet body.
On the other hand, our current observations demonstrate several close resemblance with the three-dimensional simulation results of jet-like eruptive events from \citet{{2016A&A...596A..36P}}. Their simulations suggested that the generation of jet-like eruptions in a low plasma beta condition simultaneously associate with a nonlinear torsional Alfv$\acute{e}$n wave front and a bulk of compressed plasma flows. The wave front propagates at a near local-Alven speed and is driven by the Lorentz force due to reconnection among open and closed field lines near a null point. The compressed plasma flows, namely the observed jet, are accelerated by the passage of the high-speed wave front, and then propagate along a same trajectory trailing the wave front with a lower speed. Such a spatiotemporal relationship between the Alfv$\acute{e}$n wave front and the jet is quite similar what we observed here, except that the QFP wave train ahead of the present jet should correspond to a fast-mode magnetosonic wave train. If such an identical jet generation mechanism occurs in our current case, unobservable nonlinear torsional Alfv$\acute{e}$n waves ahead of the present jet might result in the QFP wave train via some yet to be determined MHD wave mode conversion and/or amplification processes. Further observational and theoretical works are needed to validate/exclude these two possible scenarios.

In addition, some recent studies proposed that the association between type \uppercase\expandafter{\romannumeral2} radio bursts and EUV waves (or shocks) do not require similar  speeds of them, since the speed inconsistency might be attributed to different propagating directions (and/or height) in real eruptions \citep{2000A&AS..141..357K,2021ApJ...919L...7F}. However, it is still hard to associate the on-disk propagating wave train with the type \uppercase\expandafter{\romannumeral2} radio burst, since they showed not only different speeds but also different start times (see \tbl{tab1}). Besides, since the type II radio burst occurred after the jet, it might be associated with a shock in the radial direction launched during the initial stage of the jet.

\acknowledgments
The authors would like to thank the data provided by the {\em SDO}, {\em SOHO}/LASCO, {\em WIND}, {\em NRH, NDA} {\em STEREO} science teams; the anonymous referee’s valuable comments and suggestions of the paper. Y.Duan also thanks for the helpful discussions with Professor H. Tian and Dr H. Chen from Peking University. This work is supported by the Natural Science Foundation of China (12173083, 11922307, 11773068, 11633008), the Yunnan Science Foundation for Distinguished Young Scholars (202101AV070004), the Yunnan Science Foundation (2017FB006), the National Key R\&D Program of China (2019YFA0405000), the Specialized Research Fund for State Key Laboratories, and the West Light Foundation of Chinese Academy of Sciences..

 \begin{deluxetable*}{cccccccc}[b!]
\tablecaption{Relevant parameters of this event\label{tab1}}
\tablecolumns{7}
\tablenum{1}
\tablewidth{0pt}
\tablehead{
\colhead{Items} & \colhead{Start time} &
\colhead{End time} &
\colhead{During time} & \colhead{Speed} & \colhead{Height}
 \\
\colhead{ }  & \colhead{(UT)} & \colhead{(UT)} & \colhead{(min)} &
 \colhead{(km  $\rm s^{-1}$)} &  \colhead{($R_\odot$)}}
\startdata
 {EUV} Jet& 09:09 & 09:20 & 12  & 360$-$425 & \dots \\
Circular flare ribbon & 09:07 & 09:11 & 4  & \dots & \dots \\
Flare two ribbon & 09:11 & 09:20 & 9  & \dots & \dots \\
On-disk radio source & 09:08:51 & 09:09:56 & 1 & 4146$-$8414 ($\sim$ 0.0138$c-$0.028$c$) & 1.46\\
Type \uppercase\expandafter{\romannumeral3} radio burst & 09:08 & 10:00 & 52 & 6934$-$101385 (0.01$c-$0.3$c$) & \dots \\
Type \uppercase\expandafter{\romannumeral2} radio burst & 09:11 & 09:21 & 10 & 286 & 1.57$-$1.68\\
QFP wave train & 09:14 & 09:20 & 6 & 538$-$719 & \dots \\
 {EUV brightening} & 09:10 & 09:25 & 15 & 906 & \dots \\
CME(LASCO/STEREO) & 09:48/09:20 & 11:48/10:20 & 120/60 & 561/540{\tablenotemark{1}}  & \dots \\
\enddata
\tablenotetext{1}{The velocity of the CME is derived from tracing the leading edge of the CME in {\em LASCO/C2} via fitting the measured data points with a quadratic function; while the CME from STEREO-A COR1 is an average speed via tracing the leading edge of the CME.
}
\end{deluxetable*}

\end{document}